\documentclass[twocolumn]{aastex631}

\newcommand{\kms}{\rm km~s^{-1}}

\newcommand{\Msun}{{\rm M}_{\odot}}
\newcommand{\Mstar}{{\rm M}_{*}}
\newcommand{\Mdm}{{\rm M}_{\rm DM}}
\newcommand{\dn}{{\rm D}_{n}4000}
\usepackage{multirow}

\begin{document}

\title{Velocity Dispersions of Quiescent Galaxies in IllustirisTNG}

\correspondingauthor{Jubee Sohn}
\email{jubee.sohn@snu.ac.kr} 

\author[0000-0002-9254-144X]{Jubee Sohn}
\affiliation{Astronomy Program, Department of Physics and Astronomy, Seoul National University, 1 Gwanak-ro, Gwanak-gu, Seoul 08826, Republic of Korea} 
\affiliation{SNU Astronomy Research Center, Seoul National University, Seoul 08826, Republic of Korea} 

\author[0000-0002-9146-4876]{Margaret J. Geller}
\affiliation{Smithsonian Astrophysical Observatory, 60 Garden Street, Cambridge, MA 02138, USA} 

\author[0000-0002-1327-1921]{Josh Borrow}
\affiliation{Department of Physics, Kavli Institute for Astrophysics and Space Research, Massachusetts Institute of Technology, Cambridge, MA 02139, USA}

\author[0000-0001-8593-7692]{Mark Vogelsberger}
\affiliation{Department of Physics, Kavli Institute for Astrophysics and Space Research, Massachusetts Institute of Technology, Cambridge, MA 02139, USA}

\begin{abstract}
We examine the central stellar velocity dispersion of subhalos based on IllustrisTNG cosmological hydrodynamic simulations. The central velocity dispersion is a fundamental observable that links galaxies with their dark matter subhalos. We carefully explore simulated stellar velocity dispersions derived with different definitions to assess possible systematics. We explore the impact of variation in the identification of member stellar particles, the viewing axes, the velocity dispersion computation technique, and simulation resolution. None of these issues impact the velocity dispersion significantly; any systematic uncertainties are smaller than the random error. We examine the stellar mass-velocity dispersion relation as an observational test of the simulations. At fixed stellar mass, the observed velocity dispersions significantly exceed the simulation results. This discrepancy is an interesting benchmark for the IllustrisTNG simulations because the simulations are not explicitly tuned to match this relation. We demonstrate that the stellar velocity dispersion provides measures of the dark matter velocity dispersion and the dark matter subhalo mass. 
\end{abstract}

\section{INTRODUCTION} \label{sec:intro}

The nature of dark matter is a long-standing puzzle. In the $\Lambda$CDM paradigm, dark matter plays a pivotal role in the formation of galaxies and large-scale structures. Many indirect techniques trace the masses of dark matter halos. One approach is to take advantage of correlations between the properties of galaxies within dark matter (sub)halos and the dark matter halo mass. Abundance matching (e.g., \citealp{Yang03, Kravtsov04, Conroy06, Guo10, Behroozi13}), for example, compares the observed stellar mass distribution of galaxies with the dark matter mass distribution obtained from simulations. Despite its success, connecting galaxies and dark matter properties based on stellar mass is affected by potential systematic uncertainties associated with both observational and simulation-based stellar mass estimates. Stellar mass estimates may have systematic uncertainties of $\sim 0.3$ dex depending on the model choice (e.g., \citealp{Conroy06, Zahid16c}). 

Central stellar velocity dispersion is a complementary observable of galaxies that potentially traces the dark matter halo mass \citep{Wake12, Schechter15, Zahid18}. The spectroscopic stellar velocity dispersion is a more direct measure of the gravitational potential than photometric observables \citep{Wake12, Zahid18}. The typical systematic uncertainty in the stellar velocity dispersion for quiescent galaxies is also relatively small ($< 0.03$ dex, \citealp{Fabricant13, Zahid18}).

\citet{Schechter15} suggests that the stellar velocity dispersion of elliptical galaxies is a proxy for the dark matter halo mass. Based on a large sample of quiescent galaxies, \citet{Zahid16c} show that the relation between stellar velocity dispersion and the total mass is consistent with the theoretical relation for dark matter halos \citep{Evrard08, Rines16}. \citet{Utsumi20} use galaxy-galaxy weak lensing to demonstrate that the stellar velocity dispersion is proportional to the corresponding dark matter velocity dispersion (see also \citealp{vanUitert13}). 

Cosmological numerical simulations offer a testbed for investigating the relation between observables and the velocity dispersion and mass of the dark matter halo. These simulations facilitate a direct comparison between galaxy properties and dark matter halos by capturing the interplay between dark matter, stars, and gas particles within a large volume. \citet{Zahid18} use the Illustris-1 simulation \citep{Springel10, Vogelsberger14} to examine the relationship between stellar velocity dispersion and the dark matter velocity dispersion and mass. They derive the stellar velocity dispersion from simulations by mimicking fiber spectroscopy and demonstrate a tight one-to-one relation between stellar velocity dispersion and dark matter velocity dispersion.

We explore the relationship between the galaxies and their dark matter halos based on stellar velocity dispersions. We use the Illustris TNG simulations \citep{Nelson19} to extract stellar velocity dispersions for a large sample of subhalos. We derive stellar velocity dispersions following a variety of prescriptions and investigate the potential systematics. For comparison with observations, we focus on the stellar mass - velocity dispersion relation (e.g., \citealp{Zahid16c}). This comparison between observed and simulated scaling relations provides a test of IllustrisTNG because the simulations are not tuned to match the observed stellar velocity dispersions. 

We describe the IllustrisTNG simulations in Section \ref{sec:data}. We explore potential systematics in the velocity dispersion in Section \ref{sec:sigma}. We test the simulated velocity dispersion against the observed stellar mass - velocity dispersion relation in Section \ref{sec:comp_obs}. We revisit the relation between the stellar velocity dispersion and the dark matter halo in Section \ref{sec:dm}. We conclude in Section \ref{sec:conclusion}. We use the Planck cosmological parameters that are adopted for IllustrisTNG simulations with H$_{0} = 67.74~{\rm km s}^{-1}~{\rm Mpc}^{-1}$, and $\Omega_{m} = 0.3089$ and $\Omega_{\Lambda} = 0.6911$. 

\section{DATA} \label{sec:data}

We use IllustrisTNG to explore the central velocity dispersions of simulated quiescent galaxies. IllustrisTNG is a suite of cosmological magnetohydrodynamic simulations \citep{Pillepich18, Springel18, Nelson18, Naiman18, Marinacci18, Nelson19} that provides physical models of galaxy formation that aid the interpretation of observations. 

IllustrisTNG consists of three sets of simulations encompassing different volumes with differing resolution. TNG50, TNG100, TNG300 cover cubic volumes of 50 Mpc, 100 Mpc, and 300 Mpc on a side, respectively. We use TNG50-1, TNG100-1, and TNG300-1 to derive simulated velocity dispersions for statistical comparison with measurements from large spectroscopic surveys. Here, we refer to TNG50-1, TNG100-1, and TNG300-1, with the highest resolution among the simulations covering the same volumes, simply as TNG50, TNG100, and TNG300, respectively. We use only the $z = 0$ snapshot. 

We also incorporate the Illustris-1 simulation \citep{Vogelsberger14, Nelson15}, a predecessor of the TNG simulations. \citet{Zahid18} explored galaxy velocity dispersions based on Illutris-1. We compare the results from TNG and Illustris-1 in Section \ref{sec:dm}. Because the data structure of Illustris-1 is essentially identical to TNG, we can compare the results from these two simulations directly. The mass resolution of Illustris-1 is comparable with that for TNG100-1, and $\sim10$ times better than TNG300-1. 

We use subhalo catalogs based on the \texttt{SUBFIND} algorithm \citep{Springel01, Dolag09}. We first select the subhalos with total stellar mass $\Mstar > 10^{9}~\Msun$ in each of the three simulations. The mass limit corresponds roughly to the magnitude limit of dense spectroscopic surveys of local ($z < 0.1$) clusters \citep{Sohn17}. \citet{Zahid18} selected slightly less massive halos with a limiting $\Mdm > 10^{10.4}~\Msun$. They also remove subhalos with either total or stellar velocity dispersions $< 45~\kms$. 

Observations of galaxy central velocity dispersions focus on quiescent galaxies. For spiral galaxies, the interpretation of the velocity dispersion is complicated by the dominant ordered rotation. 

\begin{figure*}
\centering
\includegraphics[scale=0.35]{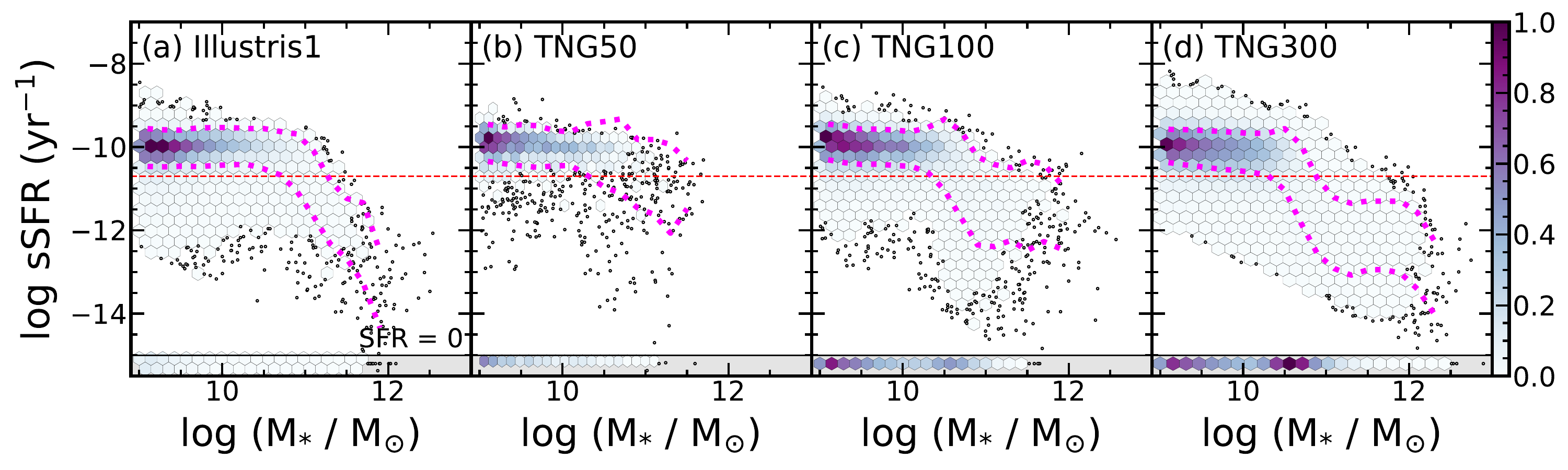}
\caption{Specific star formation rate as a function of stellar mass for subhalos in (a) Illustris-1, (b) TNG50, (c) TNG100, and (d) TNG300. Color maps show the number density: darker colors indicate higher density. Dotted lines indicate the boundaries where 68\% of the subhalos are included. The bar at the bottom of the plot, delimited by a black solid line, shows the stellar mass distribution for subhalos with SFR$ = 0$. Subhalos hosting quiescent galaxies lie below the red dashed line by selection. }
\label{fig:sfr}
\end{figure*}

We select the quiescent subhalos in simulations by using the specific star formation rate (sSFR). Here, we compute the sSFR based on the instantaneous \texttt{SFR} and the total stellar mass from the TNG \texttt{SUBFIND} catalog. Figure \ref{fig:sfr} displays the sSFR of subhalos with $\Mstar > 10^{9}~\Msun$ as a function of stellar mass. In the TNG simulations, there are many subhalos with no star formation; we assign indicative sSFRs of $-15.5$ to include these objects in the plots.  

%========================================
%========================================

We select quiescent subhalos with sSFR $< 2 \times 10^{-11}~{\rm yr}^{-1}$. This selection is much more conservative than previous choices. For example, \citet{Zahid18} select quiescent subhalos with specific star formation rate $< 2 \times 10^{-10}~{\rm yr}^{-1}$ following \citet{Wellons15}, who identified quiescent subhalos at $z = 2$. This selection identifies quiescent galaxies in Illustris-1, but it obviously includes many star-forming objects in the TNG simulations. Thus, we adopt a more conservative identification of quiescent galaxies that excludes the majority of the star-forming population. The results do not change at all with a more stringent selection, SFR$=0$. 

\begin{figure*}
\centering
\includegraphics[scale=0.35]{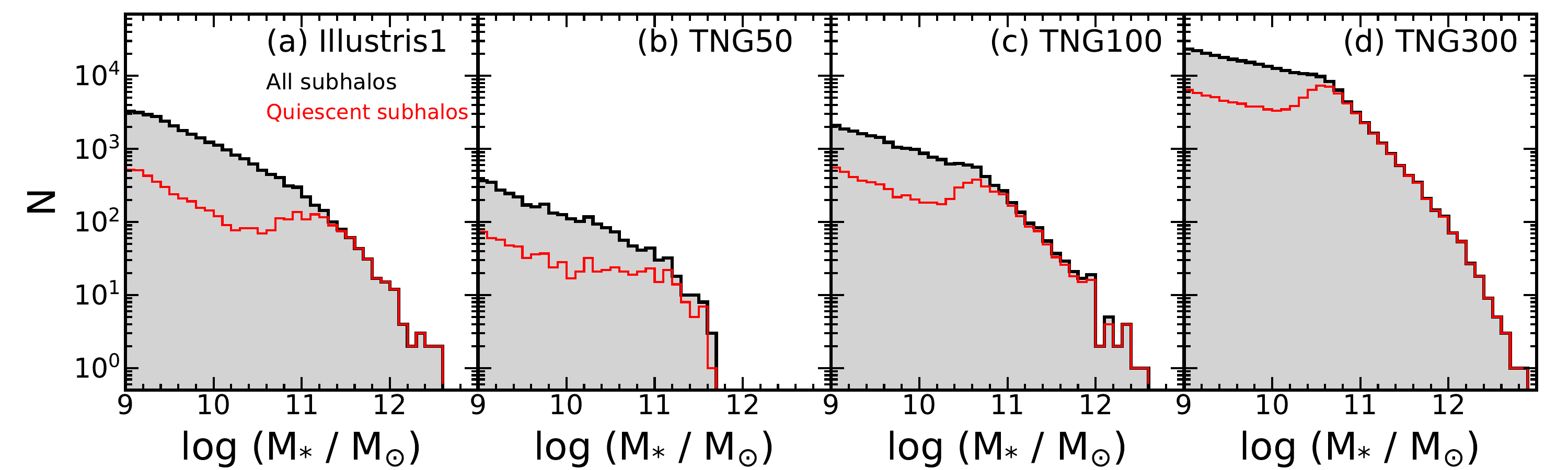}
\caption{Stellar mass distributions of subhalos in (a) Illustris-1, (b) TNG100, and (c) TNG300. Black filled histograms show the distribution of all subhalos. Red open histograms show subhalos hosting quiescent galaxies with specific star formation rate $< 2 \times 10^{-11}~{\rm yr}^{-1}$. }
\label{fig:mass_dist}
\end{figure*}

\begin{figure*}
\centering
\includegraphics[scale=0.35]{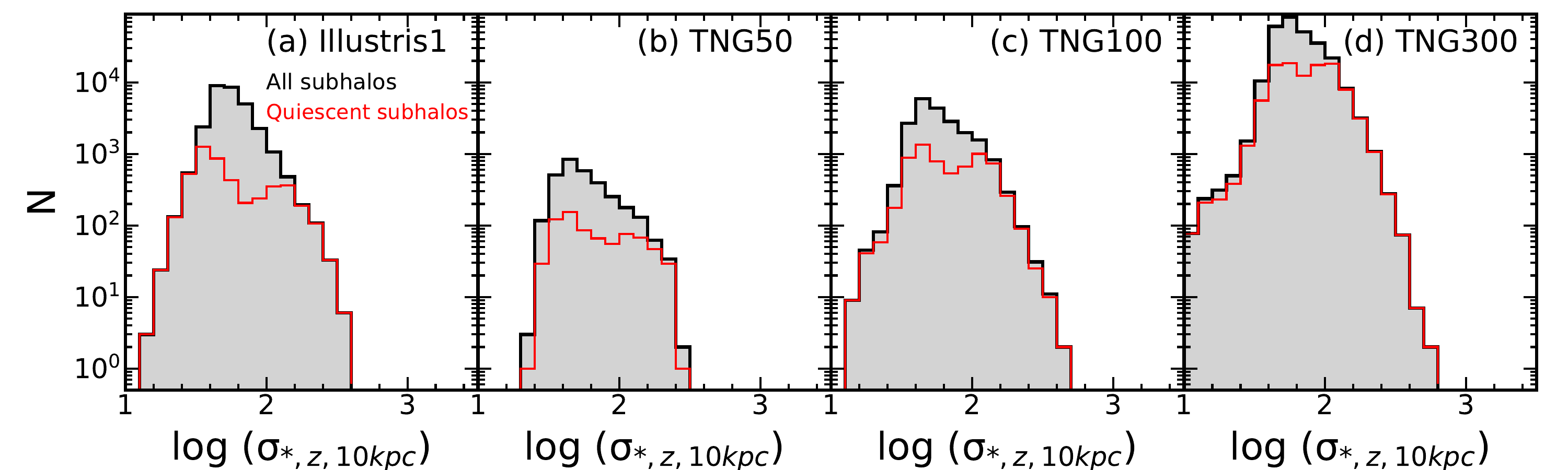}
\caption{Stellar velocity dispersion distributions of subhalos in (a) Illustris-1, (b) TNG50, (c) TNG100, and (d) TNG300. We use 1D velocity dispersions of SUBFIND member stellar particles along $z-$axis within a 10 kpc aperture based on the mass-weighted technique. Red open histograms show the same distribution of quiescent subhalos. }
\label{fig:vdisp_dist}
\end{figure*}

After selection, the samples of simulated quiescent subhalos include 4724, 734, 6619, and 104290 objects in Illutris-1, TNG50, TNG100, and TNG300, respectively. The number of subhalos in the Illustris-1 sample is smaller than the number in the sample used in \citet{Zahid18} (12,467 $z = 0$ subhalos) because of the more conservative sSFR selection. Figure \ref{fig:mass_dist} and Figure \ref{fig:vdisp_dist} display the $\Mstar$ and $\sigma_{*}$ distributions for subhalos in the three simulations. Black and red histograms display all of the subhalos and the quiescent subhalos, respectively. We note that the bump of stellar mass function at $\sim 10^{11}~\Mstar$ may result from the feedback implemented in the simulation (e.g., \citep{Salcido20}).

%========================================
%========================================
%========================================
%========================================

\section{Velocity dispersions} \label{sec:sigma}

We measure the stellar velocity dispersion of subhalos based on the particle information. The TNG \texttt{SUBFIND} catalogs include the one-dimensional (1D) velocity dispersions of all particles/cells (i.e., DM, gas, stars) associated with the subhalos. However, this velocity dispersion is not directly comparable with the observed velocity dispersion. Thus, \citet{Sohn22a} and \citet{Sohn22b} computed the velocity dispersions of the brightest cluster galaxies in TNG300 by simulating fiber spectroscopy (see also \citealp{Zahid18}). They derived the velocity dispersions within a small aperture for a fair comparison with observations. The small aperture could introduce systematics resulting from the resolution of the simulations, for example.

We test velocity dispersions based on various definitions to explore possible systematics. We first compare the 1D and 3D velocity dispersions (Section \ref{sec:sigma_dimension}). We also investigate the aperture dependence of the measurement techniques (Section \ref{sec:sigma_method}) and the velocity dispersion (Section \ref{sec:sigma_aperture}). Finally, we explore velocity dispersion measurements in simulations with different resolution (Section \ref{sec:sigma_resolution}). 

The simulated dispersion measurements we explore here have several properties in common. We obtain the spatial $x, y, z-$ center positions and central velocities of subhalos from the TNG subhalo catalogs. We base our analysis on the stellar particles associated with each subhalo identified by SUBFIND. The minimum number of stellar particles for our selected subhalos is 110; thus, the velocity dispersions are insensitive to small number statistics.

Table \ref{tab:notation} summarizes the notation we use throughout. We use $\mathrm{\sigma_{*}}$ and $\mathrm{\sigma_{DM}}$ to denote the velocity dispersion of stellar and dark matter particles, respectively. $\mathrm{\sigma_{1D}}$ and ${\mathrm{\sigma_{3D}}}$ indicate 1D and 3D velocity dispersions, respectively. We compute the 3D velocity dispersion as the quadratic sum of 1D velocity dispersion measured along the $x, y,$ and $z-$axes: ${\mathrm{\sigma_{3D}}}^{2} = (\mathrm{\sigma_{x}}^{2} + \mathrm{\sigma_{y}}^{2} + \mathrm{\sigma_{z}}^{2})$. For the measurement of 1D velocity dispersion, we consider a cylindrical volume that penetrates the subhalo. We note that the cylindrical volumes are defined separately along $x, y,$ ad $z-$axes. If the aperture for the velocity dispersion differs from our standard 10 kpc, we indicate the aperture. 

%=================================
%Table \ref{tab:notation}
%=================================
\begin{deluxetable*}{lll}
\label{tab:notation}
\tablecaption{Notation for Velocity Dispersions}
\tablecolumns{3}
\small
\tablewidth{0pt}
\tablehead{\colhead{Separation} & \colhead{Symbol} & \colhead{Description}}
\startdata
\multirow{2}{*}{Particle}   & $\mathrm{\sigma_{*}}$  &  Stellar velocity dispersion \\
                            & $\mathrm{\sigma_{DM}}$ &  Dark matter velocity dispersion \\
\hline{}
\multirow{4}{*}{Dimension}  & $\mathrm{\sigma_{1D, x}}$ & One-dimensional velocity dispersion measured along the $x-$axis \\
                            & $\mathrm{\sigma_{1D, y}}$ & One-dimensional velocity dispersion measured along the $y-$axis \\
                            & $\mathrm{\sigma_{1D, z}}$ & One-dimensional velocity dispersion measured along the $z-$axis \\
                            & $\mathrm{\sigma_{3D}}$ &  Three dimensional velocity dispersion \\
\hline{}                           
\multirow{4}{*}{Techniques} & $\mathrm{\sigma_{std}}$ & Standard deviation of velocities of particles \\
                            & $\mathrm{\sigma_{bi}}$  & Velocity dispersion measured with bi-weight technique \citep{Beers90} \\
                            & $\mathrm{\sigma_{m-weight}}$ & Mass-weighted velocity dispersion \\
                            & $\mathrm{\sigma_{l-weight}}$ & Luminosity-weighted velocity dispersion \\
\hline{}                            
\multirow{3}{*}{Aperture}   & $\mathrm{\sigma_{x kpc}}$ & Velocity dispersion measured within $x$ kpc aperture: $x = 3, 5, 10, 20, 30, 50$ kpc \\
                            & $\mathrm{\sigma_{h}}$     & Velocity dispersion measured within the half-mass radius \\
                            & $\mathrm{\sigma_{T}}$     & Velocity dispersion including all particles associated with the subhalo
\enddata 
\end{deluxetable*}
%=================================

\subsection{The Dimension of the Velocity Dispersion} \label{sec:sigma_dimension} 

We compare 3D and 1D velocity dispersions for TNG50, TNG100, and TNG300 subhalos. This comparison provides an interesting test of the impact of subhalo anisotropy. Because observations measure only the 1D line-of-sight velocity dispersions, this simulation provides useful insights that potentially connect the observed 1D velocity dispersion and the 3D velocity dispersion. Here, we measure the mass-weighted stellar velocity dispersion based on member particles within 10 kpc aperture.

\begin{figure*}
\centering
\includegraphics[scale=0.43]{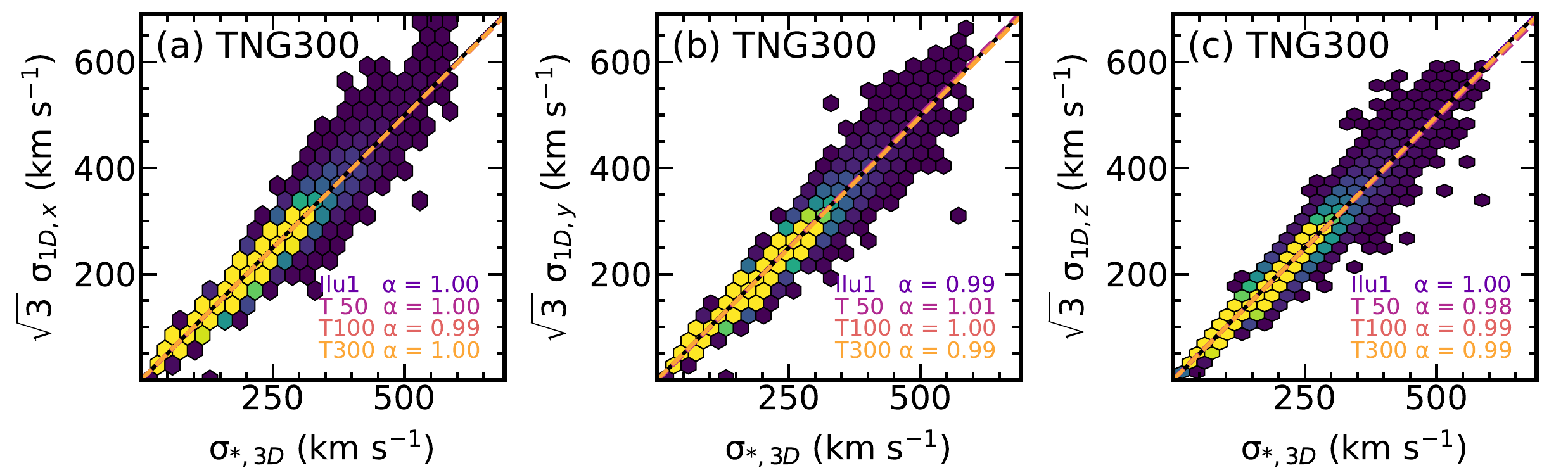}
\caption{Comparison between 3D and $\sqrt{3} \times$ 1D velocity dispersions along the $x$, $y$, and $z-$ axes (from left to right) of quiescent subhalos in TNG300. The lighter colors indicate the higher number density. Solid lines with different colors show the best-fit linear relations derived from quiescent subhalos in Illustris-1, TNG50, TNG100, and TNG300, respectively. The slope of the relations is marked in each panel.}
\label{fig:1d3d}
\end{figure*}

Figure \ref{fig:1d3d} compares $\mathrm{\sigma_{3D}}$ and $\sqrt{3} \mathrm{\sigma_{1D}}$. The density contours display the number density of the quiescent subhalos in TNG300; the lighter color indicates the higher density. From left to right, we plot 1D velocity dispersion measured along the $x-$, $y-$, and $z-$axes as a function of the 3D velocity dispersions. In general, $\mathrm{\sigma_{3D}}$ is consistent with $\sqrt{3} \mathrm{\sigma_{1D}}$. Dashed lines mark the best-fit linear relation derived from quiescent subhalos from Illustris-1, TNG 50, TNG100, and TNG 300. The slopes of the best-fit linear relations are essentially identical to one for all simulations, and regardless of the aperture we use for measuring the velocity dispersions.

%========================================
%========================================

\begin{figure*}
\centering
\includegraphics[scale=0.43]{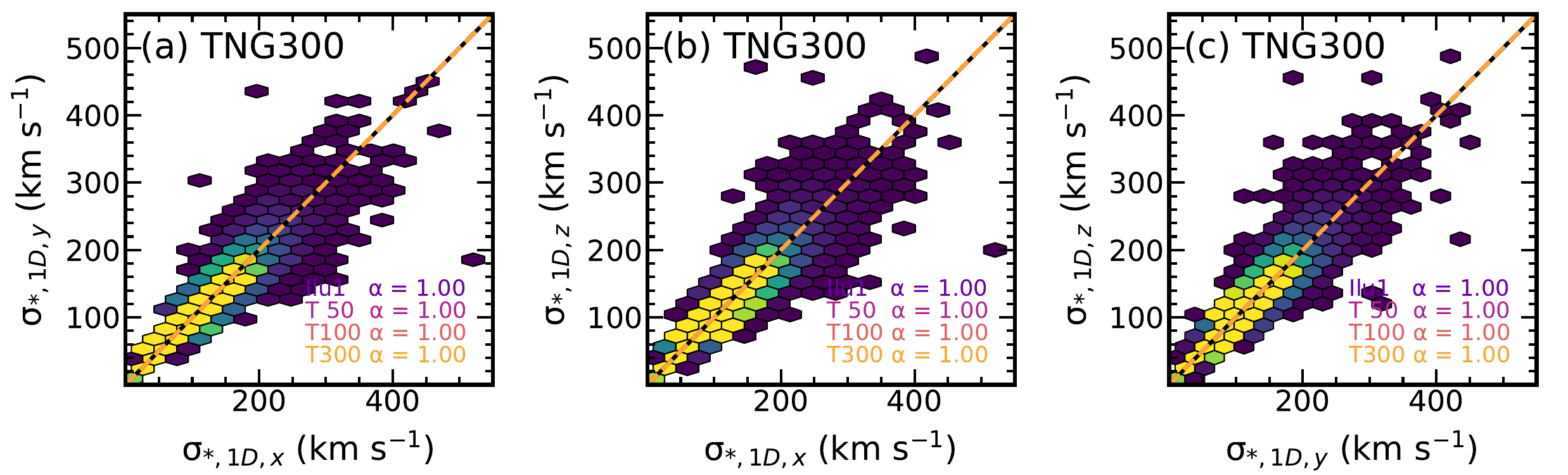}
\caption{Comparison between 1D velocity dispersion measured along (a) the $x$ and $y-$ axes, (b) $x$ and $z-$axes, and $y$ and $z-$axes of quiescent subhalos in TNG300. Color maps indicate the number density; a lighter color indicates a higher number density. Colored dashed lines mark the best-fit linear relation derived from quiescent subhalos in Illustris1, TNG50, TNG100, and TNG300. The best-fit slopes are essentially identical to unity for all simulations.}
\label{fig:axes}
\end{figure*}

We next compare 1D velocity dispersions along different axes for TNG300 subhalos (Figure \ref{fig:axes}). We use the stellar velocity dispersions based on member particles within 10 kpc using the mass-weighted technique (i.e., $\mathrm{\sigma_{*,~10 ~kpc, m-weighted}}$). From top to bottom, we show comparisons for TNG50, TNG100, and TNG300 subhalos. The left panels of Figure \ref{fig:axes} compare 1D velocity dispersions measured along the $x-$ ($\mathrm{\sigma_{*, x}}$) and $y-$axes ($\mathrm{\sigma_{*, y}}$). The middle and right panels show similar results but for $\mathrm{\sigma_{*, x}}$ versus $\mathrm{\sigma_{*, z}}$, and for $\mathrm{\sigma_{*, y}}$ versus $\mathrm{\sigma_{*, z}}$, respectively. We plot the number density of subhalos; a lighter color indicates a higher number density as in Figure \ref{fig:1d3d}. Dashed lines also show the best-fit linear relations from Illustris1, TNG50, TNG100, and TNG300. In general, the 1D velocity dispersions along different axes are consistent with one another; the best-fit relations have a slope of unity. 

%========================================
%========================================

\begin{figure*}
\centering
\includegraphics[scale=0.35]{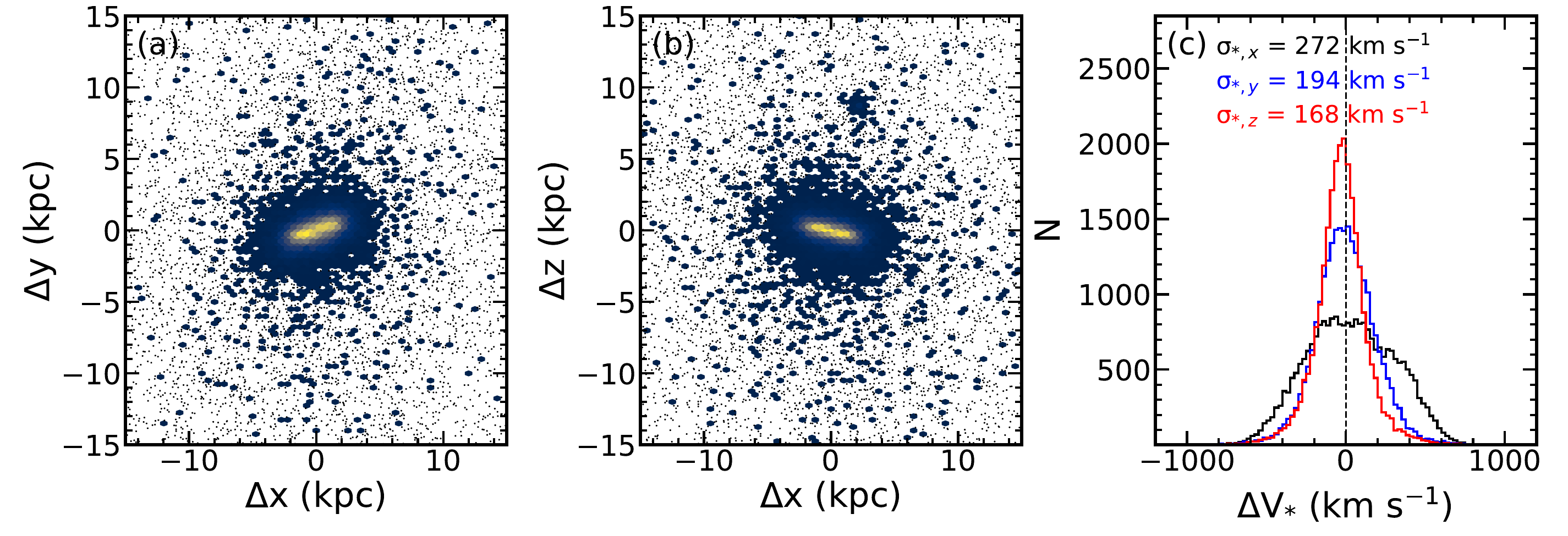} 
\caption{Spatial distributions of stars within an example halo with $\mathrm{\sigma_{*, x}}$ and $\mathrm{\sigma_{*, y}} > \mathrm{\sigma_{*, z}}$ on the (a) $x-y$ plane and (b) $x-z$ plane. (c) The relative velocity difference with respect to the subhalo center along with $x$ (black), $y$ (blue), and $z$ (red) axes. }
\label{fig:subhalo_example}
\end{figure*}

There are some individual subhalos where the 1D $\sigma_{*}$ along with a certain axis is significantly larger than others. We examine the individual subhalos with $\mathrm{\sigma_{*, z}} < \sigma_{*, x}$ or $\mathrm{\sigma_{*, z}} < \sigma_{*, y}$. The stellar particles tracing these subhalos usually have elongated distributions, indicating velocity anisotropy. Figure \ref{fig:subhalo_example} shows an example halo with with $\mathrm{\sigma_{*, z}} = 168~\kms$ and $\mathrm{\sigma_{*, x}} = 272~\kms$. The two left panels show the particle number density on the $x-y$ and $x-z$ planes. The subhalo (or the galaxy) is highly elongated in the $x-y$ and $x-z$ planes. The right panel displays the relative velocity difference between stellar particles associated with the subhalo and the mean subhalo velocity: black, blue, and red histograms show the relative velocity difference distributions along the $x-$, $y-$, and $z-$axes. The velocity distributions along with $x-$ and $y-$axes are more extended than the distribution along the $z-$axis. This example suggests that anisotropy accounts for the difference between $\mathrm{\sigma_{*, z}}$ and $\mathrm{\sigma_{*, x}}$ or $\mathrm{\sigma_{*, y}}$. The anisotropy could result from large-scale alignments, an effect that is beyond the scope of this investigation.

% %========================================
% %========================================

\subsection{Methods for Measuring the Velocity Dispersion} \label{sec:sigma_method}

We next compare velocity dispersions measured following four different approaches (see also Table \ref{tab:notation}):
\begin{itemize}
\item $\mathrm{\sigma_{std}}$: The standard deviation of the relative stellar particle velocity difference ($\Delta V_{*}$) with respect to the mean subhalo velocity. 
\item $\mathrm{\sigma_{bi}}$: The biweight velocity dispersion \citep{Beers90} based on $\Delta V_{*}$. 
\item $\mathrm{\sigma_{m-weight}}$: The mass-weighted velocity dispersion. We weight the particles according to the mass provided by TNG, and compute the standard deviation as a velocity dispersion. We specifically use `DescrStatsW' in the {\it statsmodels} package. 
\item $\mathrm{\sigma_{l-weight}}$: Similar to the mass-weighted velocity dispersion, but weighted by luminosity. We use the $g-$band luminosity of stellar particle following \citet{Zahid18}. The $g-$band covers the wavelength range ($\sim 4000 - 6000$ \AA) where the stellar velocity dispersion is measured with optical spectroscopy \citep{Thomas13, Fabricant13, Sohn20}. 
\end{itemize}
For this test, we use the 3D velocity dispersions measured based on member stellar particles within a 10 kpc aperture (i.e., $\mathrm{\sigma_{*, 3D, 10 ~kpc}}$).

\begin{figure}
\centering
\includegraphics[scale=0.26]{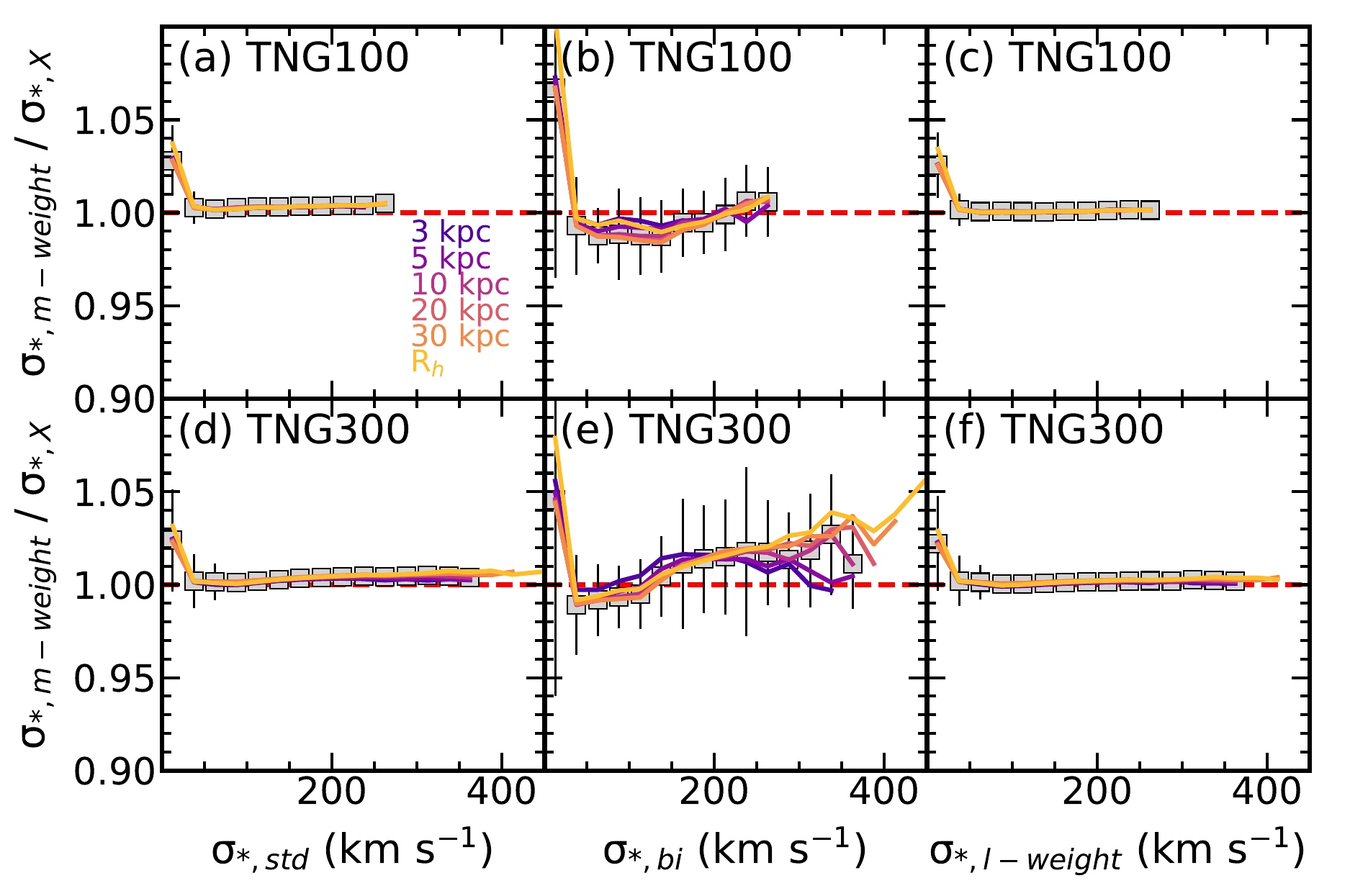}
\caption{(Upper panels) Ratio between mass-weighted velocity dispersions and other estimates: (a) the standard deviation, (b) the bi-weight velocity dispersions, and (c) the $g-$band luminosity weighted velocity dispersions. Gray squares and error bars show the median and $1\sigma$ standard deviation as a function of the various velocity dispersions measured within a 10 kpc aperture. Colored solid lines indicate the median distributions based on the velocity dispersions within various apertures. Darker colors indicate velocity dispersions within a smaller aperture. (Lower panels) Same as the upper panels, but for subhalos in TNG300. }
\label{fig:methods_ratio}
\end{figure}

Figure \ref{fig:methods_ratio} compares the mass-weighted velocity dispersion with other velocity dispersion estimates for TNG100 (upper panels) and TNG300 (lower panels) subhalos. The velocity dispersions are insensitive to the methods we use to compute them. In particular, $\mathrm{\sigma_{std}}$, $\mathrm{\sigma_{m-weight}}$, and $\mathrm{\sigma_{l-weight}}$ are essentially identical. The estimated $\mathrm{\sigma_{bi}}$s show a relatively larger scatter than other measurements, but there is little systematic difference; the median difference is much smaller than the scatter. 

%========================================
%========================================

\subsection{Aperture Effects} \label{sec:sigma_aperture}

We also compare velocity dispersions measured within various apertures. We use the 3D stellar velocity dispersions based on the mass-weighted technique (i.e., $\mathrm{\sigma_{*, 3D, m-weight}}$). 

\begin{figure}
\centering
\includegraphics[scale=0.32]{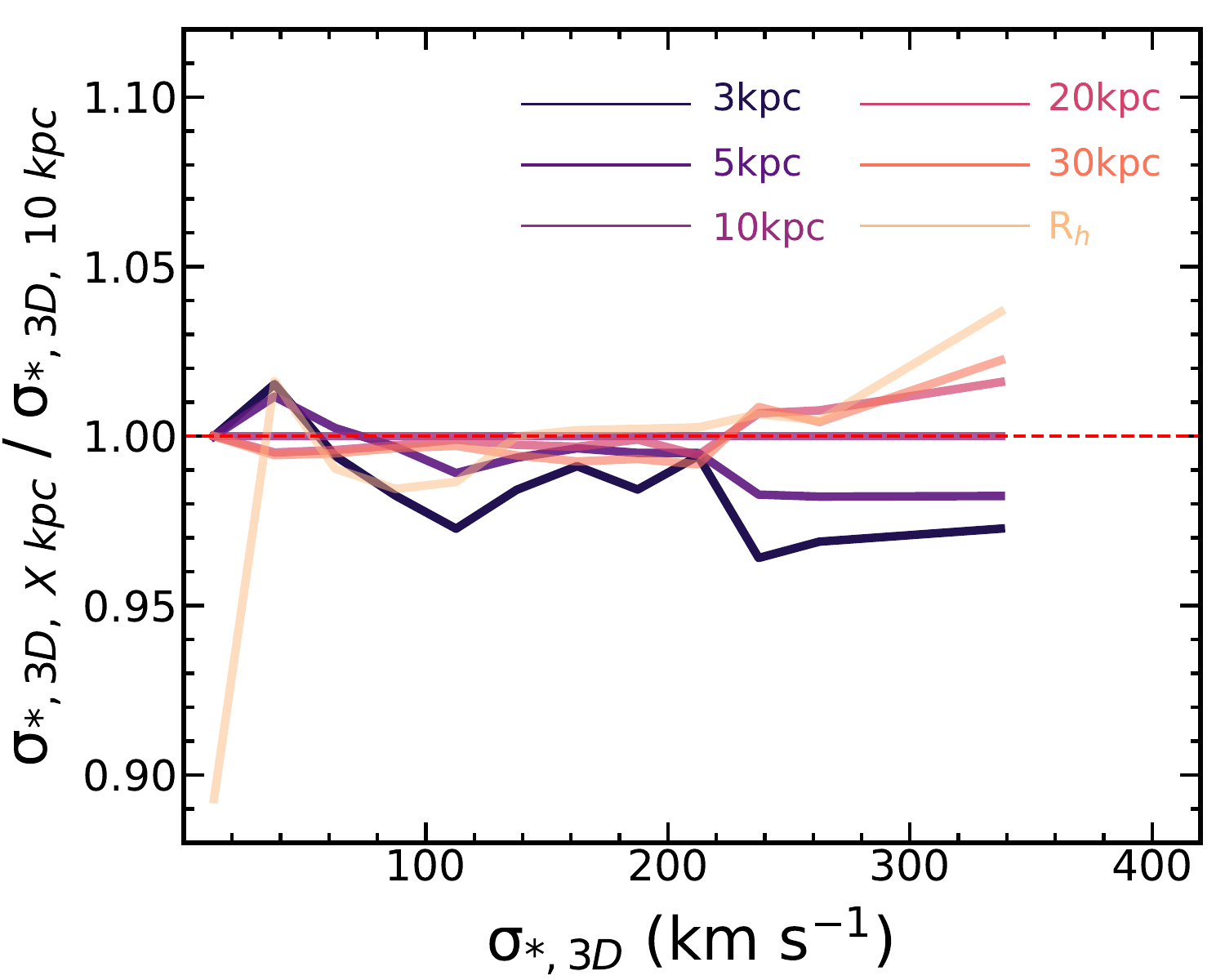}
\caption{Ratio between 3D velocity dispersion measured within various apertures with respect to the dispersion measured within 10 kpc.}
\label{fig:aperture}
\end{figure}

Figure \ref{fig:aperture} displays ratios between 3D stellar velocity dispersions measured within various apertures as a function of $\mathrm{\sigma_{*, 10~kpc}}$. We show the 3D velocity dispersions measured within 3, 5, 10, 20, and 30 kpc apertures and an aperture corresponding to $R_{h, *}$, the half-mass radius of stars. The median velocity dispersion ratios vary by $\lesssim 5\%$ depending on the aperture.

The negligible velocity dispersion dependence on the aperture indicates that the velocity dispersion profiles are generally flat within the half-mass radius. \citet{Bose21} investigate the stellar velocity dispersion profiles based on TNG300. They segregate the sample halos by mass. For halos with M$_{200} < 10^{13.5}~\Msun$, the stellar velocity dispersion profiles are flat for $R \lesssim 30$ kpc, consistent with our result. In contrast with our results, \citet{Bose21} show that the velocity dispersions of stellar particles in massive halos (M$_{200} > 10^{13.5}~\Msun$) increase by $10-25\%$ within $10 < R < 100$ kpc. The reason for this difference is unclear in part because the \citet{Bose21} method of computing the velocity dispersion is not specified in detail and may differ from our approach.

%========================================
%========================================

\subsection{Resolution} \label{sec:sigma_resolution}

Finally, we explore the impact of the resolution of the simulations on the velocity dispersion. IllustrisTNG includes simulations with differing resolutions. Depending on the resolution, the physical properties of subhalos within a DM subhalo with similar mass may differ. Furthermore, the physical scale where gravitational softening has an impact may vary.

To investigate the effect of resolution, we derive the relation between the stellar mass ($\Mstar$) of the subhalos and their stellar velocity dispersion ($\sigma_{*}$). We obtain the stellar masses of subhalos from the TNG catalog\footnote{(i.e. {\it SubhaloMassType[*,4]})}. We use the line-of-sight velocity dispersions along the $z-$axis based on the stellar particles within a 10 kpc aperture. 

\begin{figure}
\centering
\includegraphics[scale=0.26]{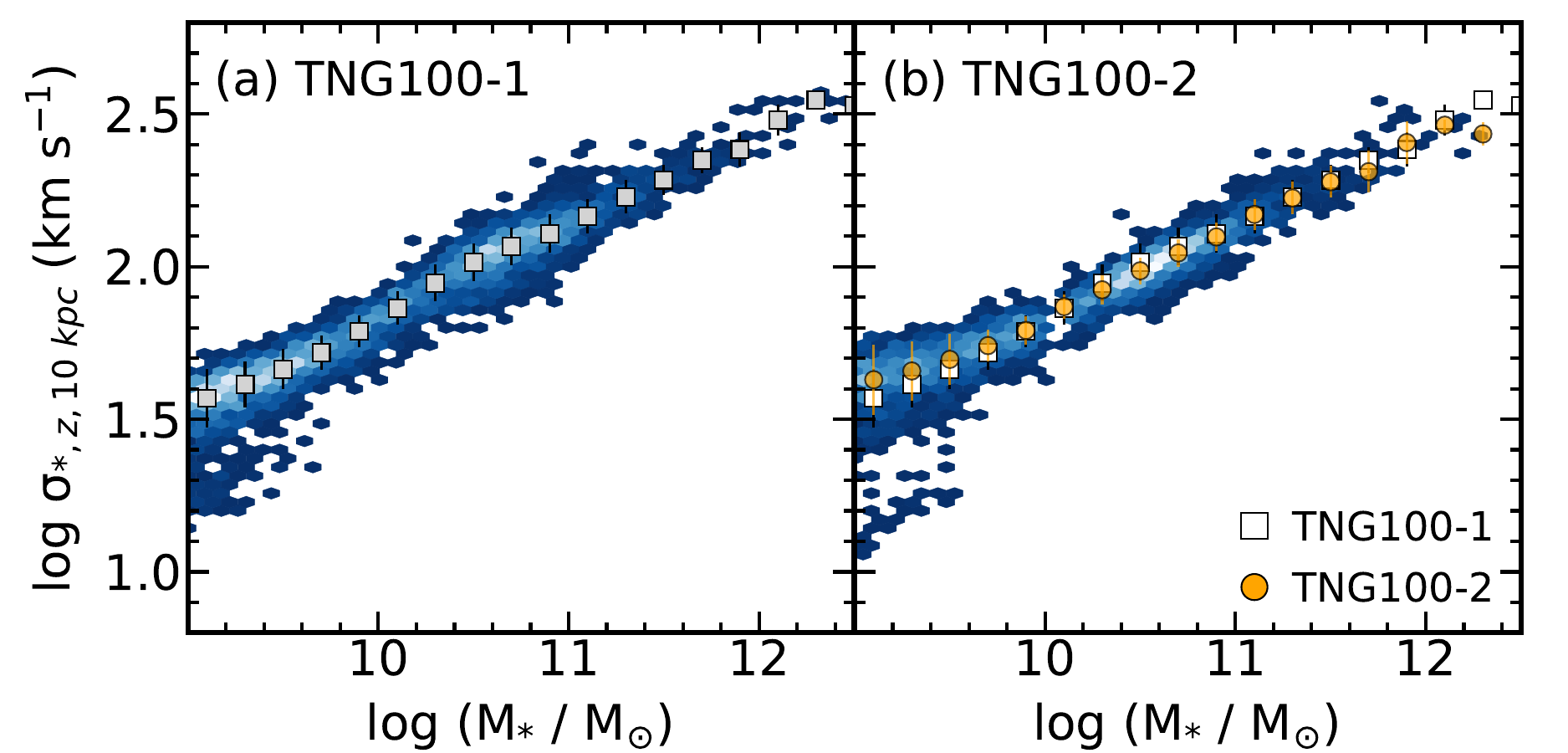}
\caption{Stellar mass vs velocity dispersion relations from (a) TNG100-1 and (b) TNG100-2. Gray squares show the median velocity dispersion as a function of stellar mass for quiescent subhalos in TNG100. Orange circles display the same relation for quiescent subhalos in TNG100-2. The error bar indicates the $1\sigma$ standard deviation in each stellar mass bin. }
\label{fig:msigma_res}
\end{figure}

Figure \ref{fig:msigma_res} compares $\Mstar - \sigma_{*}$ from TNG100-1 with TNG100-2. We use these two simulations to compare the impact of mass resolution only. The mass resolution for baryonic particles is $9.4 \times 10^{5}~\Msun / h$ and $7.6 \times 10^{6}~\Msun / h$ for TNG100-1 and TNG100-2, respectively. We analyze TNG100-2 in exactly the same manner as TNG100-1. Because of the resolution difference, the number of subhalos in the two simulations differs slightly. Nonetheless, the subhalos in TNG100-1 and TNG100-2 show very similar $\Mstar - \mathrm{\sigma_{*}}$ relations. The consistency in this relation indicates that the resolution does not significantly impact the velocity dispersion measurements for $\log~(\Mstar/\Msun) > 9.0$. We also derive the $\Mstar -\mathrm{\sigma_{*}}$ relations based on $\mathrm{\sigma_{*}}$s measured with different definitions (e.g., $\mathrm{\sigma_{*, 3D}}$ or $\mathrm{\sigma_{*}}$ and with various apertures). The relations derived from TNG100-1 and TNG100-2 are all consistent with one another regardless of the method or aperture used to compute the velocity dispersion. We also note that the $\Mstar - \mathrm{\sigma_{*}}$ relations from TNG100 and TNG300 are also consistent.

We explore various systematics in velocity dispersion measurements depending on the definition of velocity dispersion. Systematics introduced by different definitions of the velocity dispersion are at the $\sim 10\%$ level. Thus we conclude that the central velocity dispersion is a robust measure for comparing the simulations with the data and for relating the observations to the mass of the dark matter halo.

%========================================
%========================================

\section{DISCUSSION} \label{sec:discussion}

We now explore the power of the simulated central velocity dispersion as a test against observations and as a measure of the dark matter halo mass. We use an observed scaling relation as a basis for comparing the velocity dispersions from the simulations with observations (Section \ref{sec:comp_obs}). The $\Mstar - \sigma$ relation we investigate is a projection of the fundamental plane, the observed relation between size, surface brightness, and the velocity dispersion of quiescent galaxies. A projection of the fundamental plane, the $\Mstar - \sigma$ relation, provides this observational test of the simulated velocity dispersions. We also examine the relation between the stellar velocity dispersions and the dark matter subhalo velocity dispersion and mass (Section \ref{sec:dm}).

\subsection{The $\Mstar - \mathrm{\sigma_{*}}$ Relation as a Test of Illustris TNG} \label{sec:comp_obs}

The stellar mass in the $\Mstar - \mathrm{\sigma_{*}}$ is a route toward matching simulated subhalos with observed galaxies. \citet{Pillepich18} show that the stellar mass distributions based on IllustrisTNG subhalos agree well with observed stellar mass functions. 

We use the observed $\Mstar - \mathrm{\sigma_{*}}$ relation for SDSS quiescent galaxies at $0.02 < z < 0.2$ \citep{Zahid16c}. The redshift range of the SDSS sample exceeds the redshift of the simulated TNG sample at $z = 0$. However, \citet{Zahid16c} show there is negligible  redshift evolution in the $\Mstar - \mathrm{\sigma_{*}}$ relations for $z <0.6$. 

\citet{Zahid16c} select quiescent galaxies with $\dn > 1.5$, where $\dn$ is a strength of a 4000 ${\rm \AA}~$ break that correlates with the stellar population age. They also impose a mass limit, $\log~(\Mstar / \Msun) > 9$, consistent with our subhalo selection. The stellar masses of observed galaxies are based on the SDSS Model magnitudes and represent the total stellar mass (not the stellar mass estimate within a fixed aperture). The SDSS stellar masses are based on Le PHARE Spectral Energy Distribution fitting code \citep{Arnouts99, Ilbert06}. The line-of-sight velocity dispersions were obtained from \citet{Thomas13} who apply the Penalized PiXel-Fitting code (pPXF, \citealp{Cappellari04}) code to SDSS optical spectra. Because the physical coverage of the observational fixed aperture varies with redshift, \citet{Zahid16c} transform the line-of-sight velocity dispersion to a fiducial 3 kpc aperture. 

The observed quiescent galaxies in SDSS ($0.02 < z < 0.2$) define a broken power-law $\Mstar - \mathrm{\sigma_{*}}$ relation \citep{Cappellari13, Cappellari16}:
\begin{equation}
\sigma(\Mstar) = \sigma_{b} \left( \frac{\Mstar}{M_{b}} \right)^{\alpha}
\left\{
        \begin{array}{ll}
            \alpha_{1} & \text{at } \Mstar > M_{b}    \\
            \alpha_{2} & \text{at } \Mstar \leq M_{b}
        \end{array}
\right.
\end{equation}
\citet{Zahid16c} derive the best-fit power-law based on 371,884 SDSS quiescent galaxies: $\log \sigma_{b} = 2.073 \pm 0.003$, $\log (\Mstar/\Msun) = (10.26 \pm 0.01)$, $\alpha_{1} = 0.403 \pm 0.004$, and $\alpha_{2} = 0.293 \pm 0.001$. We compare this observed relation with the simulated $\Mstar - \sigma_{*}$ relation. 

\begin{figure*}
\centering
\includegraphics[scale=0.35]{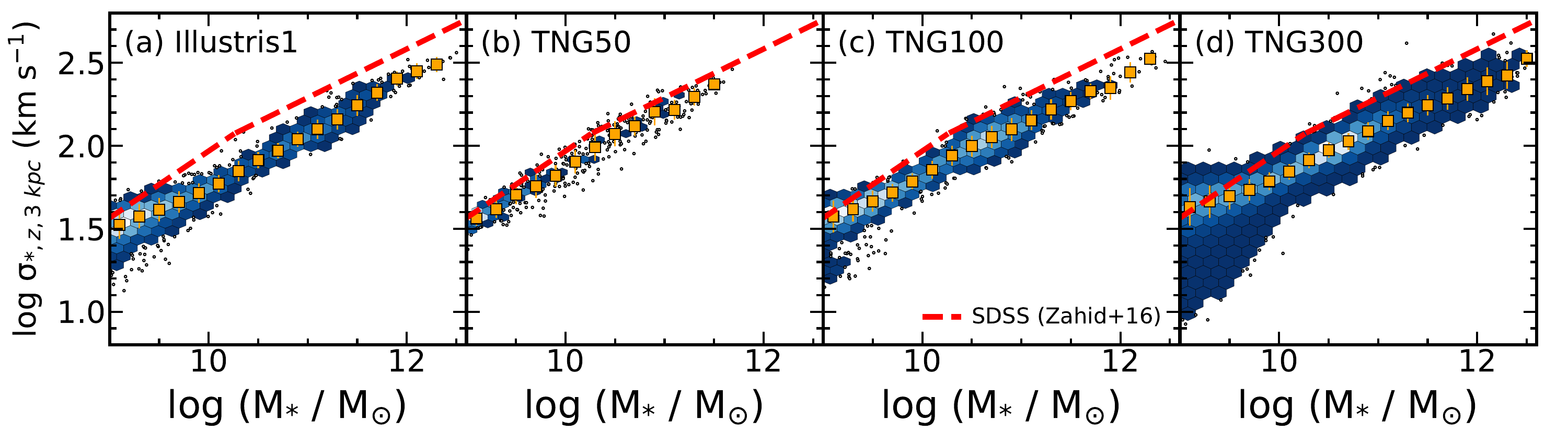}
\caption{Observed $\Mstar - \mathrm{\sigma_{*}}$ relation compared with relations for quiescent subhalos in (a) Illutris1, (b) TNG50, (c) TNG100, and (d) TNG300. Red dashed lines show the observed $\Mstar - \mathrm{\sigma_{*}}$ relation for SDSS quiescent galaxies at $0.02 < z < 0.2$ \citep{Zahid16c}. The background density maps display the distribution for the TNG subhalos. Yellow squares show the median velocity dispersions derived from the simulations. }
\label{fig:msigma_obs}
\end{figure*}

Figure \ref{fig:msigma_obs} displays the $\Mstar - \mathrm{\sigma_{*}}$ relation for  quiescent galaxies in Illustris1, TNG50, TNG100, and TNG300. We also include the Illustris1 simulations for comparison with \citet{Zahid18} who used this simulation. Red dashed lines show the relation for the SDSS quiescent galaxies. Color maps show the density distribution for simulated subhalos; a lighter color indicates higher density. The yellow squares and error bars indicate the median and the $1\sigma$ standard deviation of the distributions. Table \ref{tab:msigsim} lists the best-fit linear relation for $\Mstar - \mathrm{\sigma_{*}}$ relation for simulated subhalos. 

%=================================
%Table \ref{tab:notation}
%=================================
\begin{deluxetable*}{lcc}
\label{tab:msigsim}
\tablecaption{$\log~{\rm (\sigma_{*})} = \alpha + \beta \log~{\rm M_{*}}$}
\tablecolumns{9}
\tablewidth{0pt}
\tablehead{\colhead{Simulation} & \colhead{$\alpha$} & \colhead{$\beta$}}
\startdata
Illustris-1 & $-1.248 \pm 0.015$ & $0.302 \pm 0.001$ \\
TNG50       & $-1.555 \pm 0.036$ & $0.342 \pm 0.004$ \\
TNG100      & $-1.272 \pm 0.014$ & $0.310 \pm 0.001$ \\
TNG300      & $-0.881 \pm 0.004$ & $0.271 \pm 0.001$ 
\enddata 
\end{deluxetable*}
%=================================

The simulated velocity dispersions are generally smaller than the observed velocity dispersion at a given stellar mass, although the observed and simulated $\Mstar - \mathrm{\sigma_{*}}$ relations have a similar slope. This velocity dispersion offset appears in all of the simulations we examined. The $\Mstar - \mathrm{\sigma_{*}}$ relations based on $\mathrm{\sigma_{*, 1D, x}}$ or $\mathrm{\sigma_{*, 1D, y}}$ show the same offset.  The relation for TNG50 subhalos is slightly closer to the observed relation, but the offset remains significant. We also note that the anisotropy observed at $\sigma > 250 \kms$ when comparing $\mathrm{\sigma_{3D}}$ and $\mathrm{\sigma_{1D}}$ or contrasting $\mathrm{\sigma_{1D}}$ measurements along different axes is unrelated to this offset.

\citet{Ferrero21} also show that the $\Mstar - \mathrm{\sigma_{*}}$ relation of the subhalos in TNG100 and EAGLE simulation is offset from the observed relation. They derive the observed relation based on stellar velocity dispersions within effective radii of $\sim 300$ early-type galaxies included in ATLAS3D \citep{Cappellari13} and SLACS \citep{Bolton06} data. In their comparison, the stellar velocity dispersions of observed galaxies exceed the simulated stellar velocity dispersions at given stellar masses, consistent with our results.

%========================================
% %========================================

\subsection{Comparison between $\mathrm{\sigma_{*}}$ and Dark Matter Velocity Dispersion and Mass} \label{sec:dm}

\citet{Zahid18} show that in Illustris1, the stellar velocity dispersion is proportional to the dark matter halo velocity dispersion for both central and satellite galaxies. We test whether TNG50, TNG100, and TNG300 further support the notion that the dark matter halo mass may be inferred from the measured central velocity dispersion of quiescent galaxies.

\begin{figure*}
\centering
\includegraphics[scale=0.29]{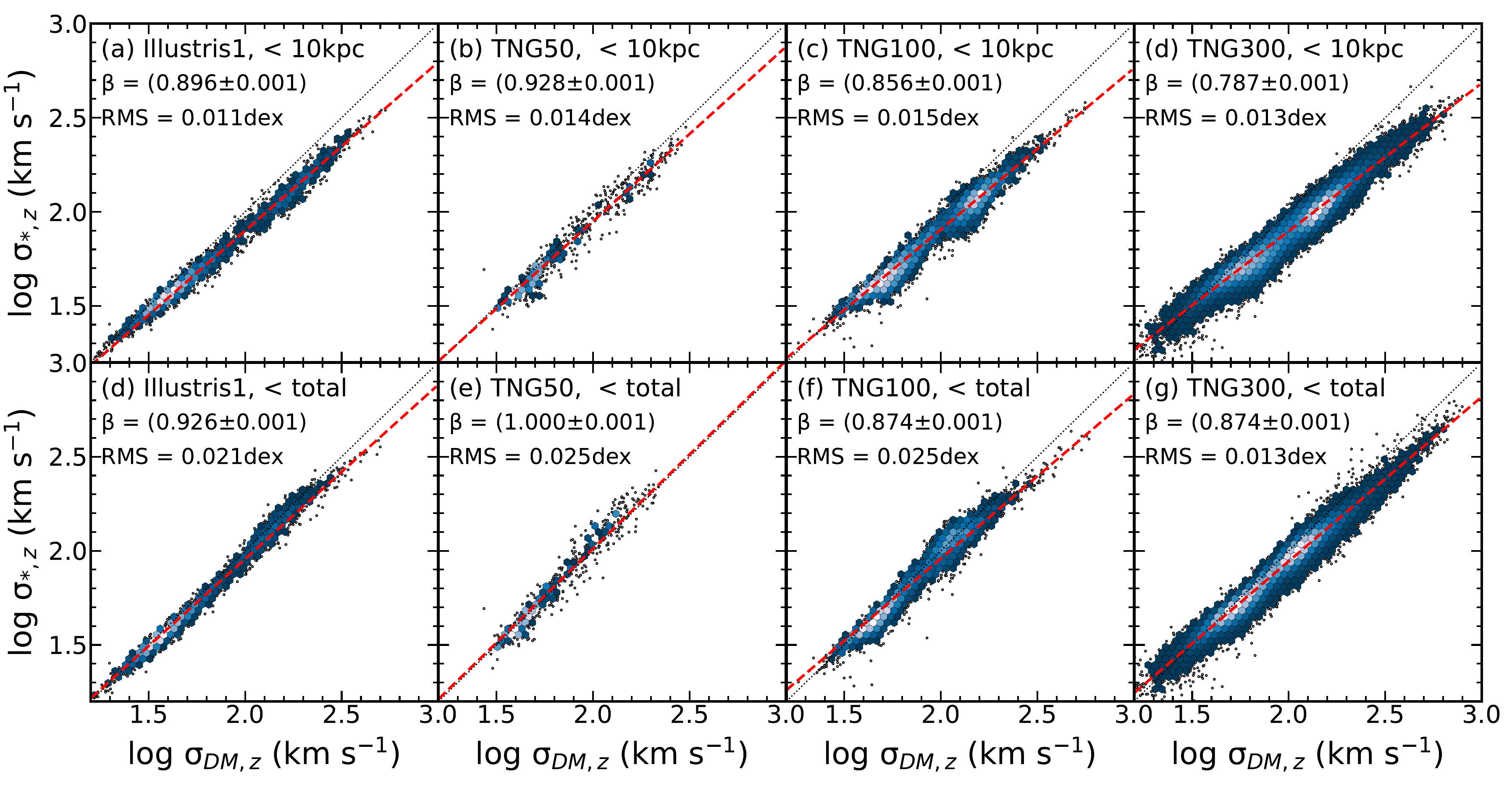}
\caption{(Upper panels) Stellar velocity dispersion as a function of Dark Matter velocity dispersion within a 10 kpc aperture for quiescent subhalos in (a) Illustris1, (b) TNG50, (c) TNG100, and (d) TNG300. Black dotted lines show the one-to-one relation, and red dashed lines mark the best-fit relation with the slope denoted in each panel. (Lower panels) Same as the upper panels, but the velocity dispersions are based on all stellar particles within each subhalo. }
\label{fig:ssig_dsig}
\end{figure*}

Figure \ref{fig:ssig_dsig} shows the $\mathrm{\sigma_{*}} - \mathrm{\sigma_{DM}}$ relations for quiescent subhalos in Illustris1, TNG50, TNG100, and TNG300. The upper panels illustrate the 3D velocity dispersions measured within a 10 kpc aperture based on the mass-weighted technique (i.e., $\mathrm{\sigma_{3D, 10~kpc, m-weighted}}$). Color maps plot the number density of the subhalos; points mark individual subhalos. Red dashed lines indicate the best-fit relations. Table \ref{tab:sigfit} summarizes the slopes and intercepts of these relations. Generally, the $\mathrm{\sigma_{*}}$s are $\sim 30-40\%$ smaller than the corresponding dark matter velocity dispersions within 10 kpc apertures. Use of 1D velocity dispersions has little impact on the slope. 

The lower panels of Figure \ref{fig:ssig_dsig} display $\mathrm{\sigma_{*}}$ and $\mathrm{\sigma_{DM}}$ for all stars and dark matter particles within the subhalos. We derive the best-fit relation based on the median $\mathrm{\sigma_{*}}$ as a function of $\mathrm{\sigma_{DM}}$. Compared to the 10 kpc velocity dispersion measurements, the total velocity dispersions agree better. In Illustris1 and TNG100, particularly, the $\mathrm{\sigma_{*}}$ and $\mathrm{\sigma_{DM}}$ are comparable with a small offset at $\mathrm{\sigma_{DM}} > 250~\kms$. This result is consistent with \citet{Zahid18}. This trend indicates that the dark matter and star particles behave more similarly at the outer region of the subhalos. 

For TNG300, the $\mathrm{\sigma_{*}}$s are $\sim 20\%$ smaller than $\mathrm{\sigma_{DM}}$ for quiescent subhalos. This result is puzzling. If the velocity dispersions reflect the total mass of the subhalos, $\mathrm{\sigma_{*}}$ and $\mathrm{\sigma_{DM}}$s should be identical. In fact, the $\mathrm{\sigma_{*}}$s are also smaller than $\mathrm{\sigma_{DM}}$ for $\mathrm{\sigma_{DM}} > 250~\kms$ in Illustris1 and TNG100. However, these offsets in Illustris1 and TNG100 are less evident because the simulation volumes are too small to include enough high-velocity dispersion subhalos. The high-velocity dispersion subhalos with $\mathrm{\sigma_{*}} > 250~\kms$ are mostly central subhalos within relatively rare massive cluster halos.

%========================================
%========================================

%=================================
%Table \ref{tab:notation}
%=================================
\begin{deluxetable*}{ccccccccc}
\label{tab:sigfit}
\tablecaption{$\log~{\rm (Y)} = \alpha + \beta \log~{\rm (X)}$}
\tablecolumns{9}
\tablewidth{0pt}
\tablehead{
\multirow{2}{*}{Y} & \multirow{2}{*}{X} & \multirow{2}{*}{Simulation} & \multicolumn{3}{c}{$< 10$ kpc} & \multicolumn{3}{c}{total} \\
%\hline{4-9}
 & & & $\alpha$ & $\beta$ & RMS (dex) & $\alpha$ & $\beta$ & RMS (dex)}
\startdata
\multirow{3}{*}{$\frac{\mathrm{\sigma_{*, x}}}{100~\kms}$} & \multirow{3}{*}{$\frac{\mathrm{\sigma_{DM, x}}}{100~\kms}$} & Illustris-1 & $0.126 \pm 0.005$ & $0.886 \pm 0.001$ & 0.010 & $0.086 \pm 0.009$ & $0.936 \pm 0.001$ & 0.017 \\
      &      & TNG50       & $0.058 \pm 0.010$ & $0.945 \pm 0.001$ & 0.013  & $-0.040 \pm 0.026$ & $1.024 \pm 0.001$ & 0.027 \\
      &      & TNG100      & $0.203 \pm 0.005$ & $0.853 \pm 0.001$ & 0.011  & $0.162 \pm 0.015$  & $0.901 \pm 0.001$ & 0.023 \\
      &      & TNG300      & $0.305 \pm 0.005$ & $0.796 \pm 0.001$ & 0.012  & $0.166 \pm 0.009$  & $0.889 \pm 0.001$ & 0.016 \\
\hline           
\multirow{3}{*}{$\frac{\mathrm{\sigma_{*, y}}}{100~\kms}$} & \multirow{3}{*}{$\frac{\mathrm{\sigma_{DM, y}}}{100~\kms}$} & Illustris-1 & $0.121 \pm 0.005$ & $0.889 \pm 0.001$ & 0.010 & $0.110 \pm 0.011$ & $0.923 \pm 0.001$ & 0.020 \\
      &      & TNG50       & $0.048 \pm 0.012$ & $0.951 \pm 0.001$ & 0.016  & $-0.095 \pm 0.022$ & $1.056 \pm 0.001$ & 0.023 \\
      &      & TNG100      & $0.192 \pm 0.006$ & $0.857 \pm 0.001$ & 0.014  & $0.202 \pm 0.013$  & $0.879 \pm 0.001$ & 0.025 \\
      &      & TNG300      & $0.333 \pm 0.005$ & $0.781 \pm 0.001$ & 0.012  & $0.189 \pm 0.006$  & $0.878 \pm 0.001$ & 0.016 \\
\hline
\multirow{3}{*}{$\frac{\mathrm{\sigma_{*, z}}}{100~\kms}$} & \multirow{3}{*}{$\frac{\mathrm{\sigma_{DM, z}}}{100~\kms}$} & Illustris-1 & $0.108 \pm  0.006$ & $0.896 \pm 0.001$ &  0.011  & $0.105 \pm 0.011 $ & $0.926 \pm  0.001$ & 0.021  \\
            &      & TNG50       & $0.093 \pm 0.011$ & $0.928 \pm 0.001$ &  0.014  & $0.013 \pm 0.024$ & $1.000 \pm 0.001$ & 0.025 \\
            &      & TNG100      & $0.193 \pm 0.007$ & $0.856 \pm 0.001$ & 0.015  & $0.211 \pm 0.012$ & $0.874 \pm 0.001$ & 0.025 \\
            &      & TNG300      & $0.322 \pm 0.005$ & $0.787 \pm 0.001$ & 0.013  & $0.196 \pm 0.005$ & $0.874 \pm 0.001$ & 0.013 \\
\hline
\multirow{3}{*}{$\frac{\mathrm{\sigma_{*, 3D}}}{100~\kms}$} & \multirow{3}{*}{$\frac{\mathrm{\sigma_{DM, 3D}}}{100~\kms}$} & Illustris-1 & $0.215 \pm 0.004$ & $0.861 \pm 0.001$ &  0.011  & $0.117 \pm 0.007$ & $0.930 \pm 0.001$ & 0.017 \\
             &      & TNG50       & $0.118 \pm 0.013$ & $0.924 \pm 0.001$ & 0.016  & $-0.091 \pm 0.028$ & $1.050 \pm 0.001$ & 0.026 \\ 
             &      & TNG100      & $0.258 \pm 0.006$ & $0.846 \pm 0.001$ & 0.014  & $0.219 \pm 0.009$ & $0.885 \pm 0.001$ & 0.022 \\
             &      & TNG300      & $0.371 \pm 0.004$ & $0.788 \pm 0.001$ & 0.012  & $0.255 \pm 0.004$ & $0.861 \pm 0.001$ & 0.010
\enddata 
\end{deluxetable*}
%=================================

\begin{figure*}
\centering
\includegraphics[scale=0.29]{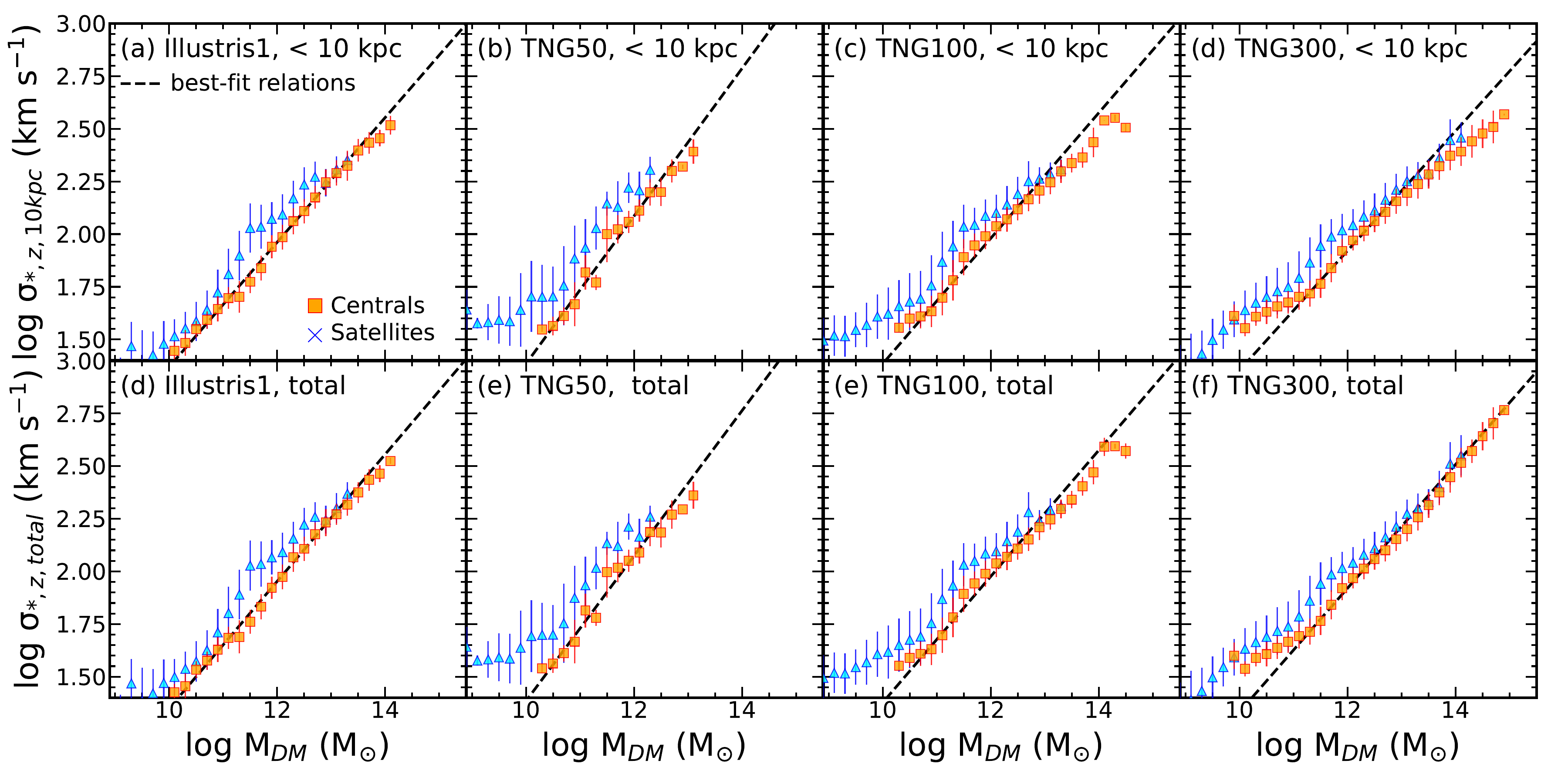}
\caption{(Upper panels) Stellar velocity dispersions measured along with $z-$axis based on stellar particles within a 10 kpc aperture as a function of the dark matter mass of subhalos ($\Mdm$) in (a) Illustris1, (b) TNG50 (c) TNG100, and (d) TNG300. Red squares and blue triangles show the median velocity dispersion as a function of $\Mdm$ of central and satellite subhalos, respectively. Black dashed lines show the best-fit relation derived from central subhalos. (Lower panels) Same as upper panels, but for total stellar velocity dispersion based on all stellar particles associated with each subhalo. }
\label{fig:sigma_mdm}
\end{figure*}

We next derive the relations between $\mathrm{\sigma_{*}}$ and the dark matter mass of subhalos. Figure \ref{fig:sigma_mdm} illustrates $\Mdm - \mathrm{\sigma_{*, z}}$ relations derived from (leftmost) Illustris1, (middel-left) TNG50, (middle-right) TNG100, and (rightmost) TNG300. Here, $\Mdm$ indicates the sum of dark matter particles associated with the subhalos. The upper panels show $\mathrm{\sigma_{*, z}}$ based on stellar particles within 10 kpc from the subhalo center, and the lower panels display $\mathrm{\sigma_{*, z}}$ using all stellar particles within the subhalos. $\mathrm{\sigma_{*, z, 10 kpc}}$ is directly measurable, while $\mathrm{\sigma_{*, z, total}}$ is challenging to derive from observations. Thus, the upper panels elucidate whether we can infer the underlying dark matter subhalo mass based on the observable velocity dispersions. The lower panels confirm the general scaling relation between $\mathrm{\sigma_{*}}$ and $\Mdm$. 

We segregate the central and satellite subhalos. We identify subhalos that are primary subhalos\footnote{selected based on \texttt{GroupFirstSub} in the \texttt{SUBFIND} catalog} within group/cluster halos in the Friends-of-Friends catalogs of the simulations. In Figure \ref{fig:sigma_mdm}, we plot the median $\mathrm{\sigma_{*, z}}$s as a function of $\Mdm$ for central subhalos (red squares) and satellites subhalos (blue triangles), respectively. Dashed lines show the best-fit relations for central subhalos. Table \ref{tab:msigfit} summarizes the best-fit relation parameters. In general, the dark matter mass is proportional to $\mathrm{\sigma_{*, z}}^{3.3 - 3.5}$. 

Central subhalos consistently follow the $\Mdm \propto \sigma_{, z}^{3.3}$ relation across a wide range of $\Mdm$, particularly in Illustris1, TNG50, and TNG100. However, TNG300 central subhalos with $\Mdm > 10^{13.5}~\Msun$ exhibit lower $\mathrm{\sigma_{, z}}$ values than expected based on the $\Mdm \propto \mathrm{\sigma_{, z}}^{3.3}$ relation. A few TNG100 subhalos also show similar offsets in $\mathrm{\sigma_{*}}$. This discrepancy probably arises from the aperture we use to measure $\mathrm{\sigma_{, z}}$ because there is no offset in the $\Mdm - \mathrm{\sigma_{*, z, total}}$ relation.

Central subhalos also follow the $\Mdm \propto \mathrm{\sigma_{*, z}}^{3.3}$ relation over a wide range of $\Mdm$, particularly in Illustris1, TNG50, and TNG100. However, TNG300 central subhalos with $\Mdm > 10^{13.5}~\Msun$ exhibit lower $\mathrm{\sigma_{*, z}}$ values than predicted from the $\Mdm \propto \mathrm{\sigma_{*, z}}^{3.3}$ relation. A few TNG100 subhalos show similar offsets in $\sigma_{*}$. The discrepancy results from the aperture for measuring $\mathrm{\sigma_{*, z}}$ because there is no offset in the $\Mdm - \mathrm{\sigma_{*, z, total}}$ relation. 

Satellite subhalos generally exhibit larger $\mathrm{\sigma_{*, z}}$ than predicted for $\Mdm$ with $\Mdm < 10^{12}~\Msun$. The presence of his offset in the $\Mdm - \mathrm{\sigma_{*, z, total}}$ relation indicates that it is not an aperture effect. \citet{Zahid18} find a similar offset in the Illustris1 simulation and suggest that the difference results from tidal stripping of dark matter subhalos during infall onto cluster halos. We confirm the difference they find with a larger sample extracted from larger simulations. The tight $\Mdm - \mathrm{\sigma_{*, z, 10~kpc}}$ relation suggests that $\mathrm{\sigma_{*, z}}$ is a measure of $\Mdm$. For $\mathrm{\sigma_{*}} > 125~\kms$ the estimate of $\Mdm$ is reasonable because both centrals and satellites share a similar $\Mdm - \mathrm{\sigma_{, z}}$ relation within this $\mathrm{\sigma_{*}}$ range. 

%========================================
%========================================

%=================================
%Table \ref{tab:notation}
%=================================
\begin{deluxetable*}{cccccccc}
\label{tab:msigfit}
\tablecaption{$\log~{\rm \Mdm} = \alpha + \beta \log~{\mathrm{\sigma_{*, z, 10~kpc}}}$}
\tablecolumns{8}
\tablewidth{0pt}
\tablehead{
\multirow{2}{*}{Simulation} & \multirow{2}{*}{Subhalos} & \multicolumn{3}{c}{$< 10$ kpc} & \multicolumn{3}{c}{total} \\
 & & $\alpha$ & $\beta$ & RMS (dex) & $\alpha$ & $\beta$ & RMS (dex)}
\startdata
\multirow{4}{*}{All Subhalos} & Illustris1 & $-0.124 \pm 0.008 $ & $3.280  \pm 0.020$ & 0.307 & $-0.101 \pm 0.008$ & $3.226 \pm 0.020$ & 0.312 \\ 
                              & TNG50      & $0.154  \pm 0.007 $ & $3.343 \pm 0.024$ & 0.191 & $0.180  \pm 0.007$ & $3.281 \pm 0.023 $ & 0.187 \\ 
                              & TNG100     & $-0.422 \pm 0.010$ & $2.663 \pm 0.024$ & 0.251 & $-0.405 \pm 0.010$ & $ 2.615  \pm  0.024 $ &  0.259 \\ 
                              & TNG300     & $-0.693 \pm 0.023$ & $3.022 \pm 0.077$ & 0.464 & $-0.668 \pm 0.024$ & $ 3.057  \pm  0.080 $ &  0.473 \\
\hline                              
\multirow{4}{*}{Centrals}     & Illustris1 & $-0.210  \pm  0.018 $ & $ 2.863  \pm  0.073 $ &  0.207  & $-0.180  \pm  0.019 $ & $ 2.911  \pm  0.079 $ &  0.220 \\ 
                              & TNG50      & $-0.967  \pm  0.027 $ & $ 2.400  \pm  0.086 $ &  0.402  & $-0.952  \pm  0.028 $ & $ 2.412  \pm  0.089 $ &  0.409 \\
                              & TNG100     & $-0.260  \pm  0.008 $ & $ 3.666  \pm  0.028 $ &  0.434  & $-0.247  \pm  0.008 $ & $ 3.662  \pm  0.028 $ &  0.436 \\ 
                              & TNG300     & $0.096  \pm  0.006 $ & $ 3.387  \pm  0.032 $ &  0.232  & $0.109  \pm  0.006 $ & $ 3.369  \pm  0.032 $ &  0.234 \\
\hline
\multirow{4}{*}{Satellites}   & Illustris1 & $-0.615  \pm  0.010 $ & $ 2.821  \pm  0.033 $ &  0.372  & $-0.605  \pm  0.011 $ & $ 2.818  \pm  0.034 $ &  0.375 \\
                              & TNG50      & $-0.009  \pm  0.002 $ & $ 4.023  \pm  0.008 $ &  0.428  & $-0.005  \pm  0.002 $ & $ 3.944  \pm  0.007 $ &  0.420 \\ 
                              & TNG100     & $0.306  \pm  0.001 $ & $ 3.527  \pm  0.007 $ &  0.223  & $0.301  \pm  0.001 $ & $ 3.414  \pm  0.007 $ &  0.207 \\ 
                              & TNG300     & $-0.389  \pm  0.002 $ & $ 3.168  \pm  0.009 $ &  0.359  & $-0.376  \pm  0.002 $ & $ 3.137  \pm  0.009 $ &  0.358 \\
\enddata 
\end{deluxetable*}
%=================================

\section{CONCLUSION} \label{sec:conclusion}

We investigate the stellar velocity dispersion of galaxy subhalos in IllustrisTNG as a fundamental observable tracer of the underlying dark matter subhalos. We first explore possible systematics in measuring the velocity dispersion. We compute the velocity dispersion by varying the identification of the member stellar particles, the viewing axes, the velocity dispersion calculation technique, and the simulation resolution. In summary, the velocity dispersions are insensitive to all of these issues. The scatter in the velocity dispersion distribution is generally larger than the systematic uncertainty.

Comparison with observations offers an interesting test of the models. The IllustrisTNG simulations are not tuned to match the $\Mstar - \mathrm{\sigma_{*}}$ relation. We thus compare the $\Mstar - \mathrm{\sigma_{*}}$ relations from simulations with SDSS at $z < 0.2$. The observed $\mathrm{\sigma_{*}}$ are generally larger than the simulated $\mathrm{\sigma_{*}}$ by $\sim 40-50\%$ at a given stellar mass. Explaining the observed velocity dispersion offset may provide constraints on the feedback models implemented in the simulations. 

We compare the stellar velocity dispersions of the stellar and dark matter particles: $\mathrm{\sigma_{*}}$ and $\mathrm{\sigma_{DM}}$. The stellar velocity dispersions show a one-to-one correlation with the dark matter velocity dispersion for $\mathrm{\sigma_{DM}} < 400~\kms$. The stellar velocity dispersions are smaller than $\mathrm{\sigma_{DM}}$ for $\mathrm{\sigma_{DM}} > 400~\kms$. This offset is particularly pronounced in TNG300, where the massive subhalos are more abundant. 

The $\Mdm - \mathrm{\sigma_{*}}$ relation tests $\mathrm{\sigma_{*}}$ as a tracer of the underlying dark matter subhalo mass. Following \citet{Zahid18}, we derive the $\Mdm - \mathrm{\sigma_{*}}$ relations for central and satellite subhalos separately. For central subhalos, $\mathrm{\sigma_{*}}$ is tightly correlated with $\Mdm$: $\Mdm \propto \mathrm{\sigma_{*}}^{3.3}$. For satellite subhalos, the $\Mdm - \mathrm{\sigma_{*}}$ relation is slightly shallower than the relation for central subhalos: $\Mdm \propto \mathrm{\sigma_{*}}^{3.0}$, consistent with  \citet{Zahid18} who attribute the difference to tidal stripping of the satellite dark matter subhalos. 

Upcoming cosmological simulations covering a larger volume (e.g., MillenniumTNG \citep{Pakmor23}, The Flamingo simulations \citep{Schaye23}) will deliver a huge sample of velocity dispersions, including many subhalos in massive cluster halos. Furthermore,  future wide-field multi-object spectroscopic surveys (e.g., 4MOST \citep{deJong19}, MOONS \citep{Maiolino20}, Subaru PFS \citep{Takada14}) will provide an extremely large sample of velocity dispersion based on deep and dense spectroscopy. Comparison between the velocity dispersions from future simulations and spectroscopy will enable more stringent tests of galaxy formation recipes implemented in simulations. Combining these velocity dispersions with other photometric observables obtained from high-resolution imaging surveys (e.g., EUCLID, LSST) will enable galaxy mass estimates that combine weak lensing with spectroscopic observables to deepen understanding of the relationship between galaxies and their dark matter halos.

\begin{acknowledgments}
We thank Scott Kenyon, Ivana Damjanov, Antonaldo Diaferio, Ken Rines, and Michele Pizzardo for insightful comments that clarify this manuscript. This work was supported by the National Research Foundation of Korea(NRF) grant funded by the Korea government(MSIT) (RS-2023-00210597).
\end{acknowledgments}

\vspace{5mm}
%\facilities{HST(STIS), Swift(XRT and UVOT), AAVSO, CTIO:1.3m, CTIO:1.5m,CXO}

\software{astropy \citep{astropy13, astropy18}}

\bibliography{ms}{}
\bibliographystyle{aasjournal}

\end{document}